# Perfect topological charge for asymptotically free theories [*]

M. Blatter, R. Burkhalter, P. Hasenfratz and F. Niedermayer[a]

[a]Institute for Theoretical Physics, University of Bern,
Sidlerstrasse 5, CH-3012 Bern, Switzerland

The classical equations of motion of the perfect lattice action in asymptotically free $d=2$ spin and $d=4$ gauge models possess scale invariant instanton solutions. This property allows the definition of a topological charge on the lattice which is perfect in the sense that no topological defects exist.

The basic construction is illustrated in the $d=2$ O(3) non–linear $\sigma$–model and the topological susceptibility is measured to high precision in the range of correlation lengths $\xi \in (2-60)$. Our results strongly suggest that the topological susceptibility is not a physical quantity in this model.

## 1. Introduction

Classical topological solutions (instantons) might play an important role in nonperturbative dynamics of asymptotically free quantum field theories. Standard lattice discretization, however, breaks scale invariance and the corresponding classical equations have no stable instanton solutions. In addition, at least in the geometric definition of the topological charge, there exist topological defects, i.e. configurations which carry topological charge but with an action much below the continuum value. All these problems slowed down progress in this field [1]. In this paper we discuss how one can avoid these problems and we present numerical results of the topological susceptibility in the $d=2$ O(3) non–linear $\sigma$–model.

This paper strongly builds on earlier papers by two of us [2]. For a more detailed discussion on some of the equations and notions in section 2 we refer the reader to these references.

## 2. Perfect lattice action in the O(3) non–linear $\sigma$–model

Consider a square lattice with the spins $\mathbf{S}_n$ ($\mathbf{S}_n^2 = 1$) sitting at the lattice sites. The lattice is divided into $2 \times 2$ blocks and to every block we associate a block spin $\mathbf{R}_{n_B}$. The perfect classical action $\mathcal{A}^*$ is the Fixed Point (FP) of the Renormalization Group (RG) transformation. It is determined by the implicit FP equation

$$\mathcal{A}^*(\mathbf{R}) = \min_{\{\mathbf{S}\}} \{T(\mathbf{R}, \mathbf{S}) + \mathcal{A}^*(\mathbf{S})\}, \qquad (1)$$

where

$$T(\mathbf{R}, \mathbf{S}) = \kappa \sum_{n_B} \left[ |\sum_{n \in n_B} \mathbf{S}_n| - \mathbf{R}_{n_B} \sum_{n \in n_B} \mathbf{S}_n \right]. \quad (2)$$

Here $\kappa$ is a free parameter that is used to optimize the corresponding FP action.

We may solve this equation iteratively, i.e. we introduce a multigrid with the input configuration $\{\mathbf{R}\}$ on the coarsest level. On the finest level we can use any lattice action, e.g. the standard action.

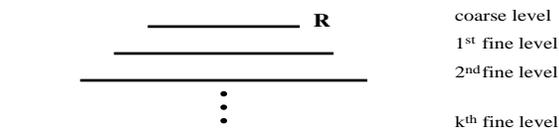

Figure 1. Multigrid for minimizing the FP equation

A parametrization of $\mathcal{A}^*$ can be obtained using 24 couplings (which involve 2–, 3– and 4–spin couplings). We found a parametrization which gives even for coarse configurations and notably for classical instanton configurations only negligible deviations from the minimized values.

---

[*]Presented by R. Burkhalter

## 3. Classical solutions, instantons and the perfect topological charge

Some important statements can be made for $\mathcal{A}^*$ from equation 1 even without solving it explicitly. For instance the

**Statement**:
If the configuration $\{\mathbf{R}\}$ satisfies the FP classical equations and it is a local minimum of $\mathcal{A}^*(\mathbf{R})$, then the configuration $\{\mathbf{S}(\mathbf{R})\}$ on the finer lattice which minimizes the right hand side of eq. 1 satisfies the FP equations as well. In addition, the value of the action remains unchanged: $\mathcal{A}^*(\mathbf{S}(\mathbf{R})) = \mathcal{A}^*(\mathbf{R})$.

This implies for the O(3) non-linear $\sigma$-model that, if $\mathcal{A}^*$ has some instanton solution of size $\rho$, then there exist instanton solutions of size $2\rho$, $4\rho$, ... The converse statement, however, need not be true. We construct instanton solutions by exploiting the above statement: We put on a very fine lattice the exact 2-instanton continuum solution on a torus [3]. The radii of the instantons are $\rho \cdot 2^k$ with $\rho$ of order one lattice spacing. Then we perform $k$ block-transformations:

$$\mathbf{R}_{n_B} = \sum_{n \in n_B} \mathbf{S}_n \, / \, |\sum_{n \in n_B} \mathbf{S}_n| \qquad (3)$$

so that the final instantons have radii of order $a$. For these configurations we measure the exact $\mathcal{A}^*$ on the multigrid by minimization, the parametrized FP action and the topological charge.

If we use the geometric definition of the topological charge [4], it is possible to have configurations with $Q = 2$ even if $\mathcal{A}^* \leq 8\pi$ (in contradiction to continuum). This means that topological defects will still be present. Hence we should not only define a perfect lattice action but also a perfect lattice topological charge. We may do this by calculating $Q$ not on the given configuration but on a finer level in the minimizing multigrid using the geometric definition. This makes sense because in the (albeit infinite) multigrid the finest level configuration obtained after minimizing the iterated FP equation is the continuum equivalent of the coarse input configuration. This definition is perfect in the sense that always $\mathcal{A}^* \geq 4\pi|Q|$ and $\mathcal{A}^* = 4\pi|Q|$ for instanton solutions (independently of $\rho$), i.e. there are no topological defects.

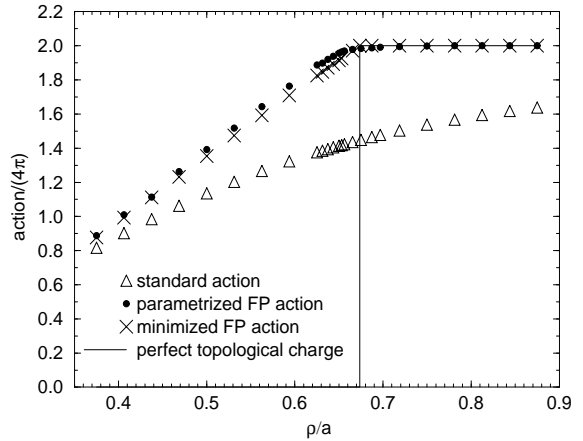

Figure 2. Action and topological charge of smallest possible 2-instanton solutions on the lattice

Figure 2 shows the different actions and charges of configurations with $\rho$ of order $a$: Below $\rho \lesssim 0.7a$ the instantons have disappeared. In the region where the instantons still exist, $\mathcal{A}^*$ has the correct continuum value $8\pi$. It is evident from Fig. 2 that the parametrization for the FP action is very accurate whereas the standard action is quite bad.

In order to reduce calculational effort, a parametrization of the minimizing fine configuration is needed. For such a parametrization we take the following ansatz

$$\mathbf{S}_n = \sum_{n_B} \alpha(n, n_B) \mathbf{R}_{n_B} \qquad (4)$$
$$+ \sum_{\substack{n_B \\ m_B, m'_B}} \beta(n, n_B; m_B, m'_B) \mathbf{R}_{n_B} (1 - \mathbf{R}_{m_B} \mathbf{R}_{m'_B}).$$

Similar to the situation for the perfect action, we can calculate analytically the coefficients $\alpha$, the coefficients $\beta$ we determine by fitting.

## 4. Topological Susceptibility

If the topological susceptibility

$$\chi_t = <Q^2>/V \qquad (5)$$

is a physical quantity, one expects that $\chi_t \cdot \xi^2$ is constant in the continuum limit $\xi \to \infty$. If cut-off effects were present, which may occur due to the lower bound in instanton sizes, then the continuum limit is reached with $\xi^{-2}$-corrections. For earlier numerical works on the topological susceptibility in this model and further references we refer the reader to [4,5].

We measure the topological susceptibility with a Wolff-type cluster algorithm at correlation lengths in the range $\xi \in (2-60)$ using the perfect lattice action and the perfect charge. For the perfect charge we use the parametrization of the fine field and measure the charge on the first finer level. In order to avoid finite size effects we keep the ratio $L/\xi \approx 6$ constant. The results are shown in Figure 3. Clearly, no scaling is seen even at correlation lengths as large as 60.

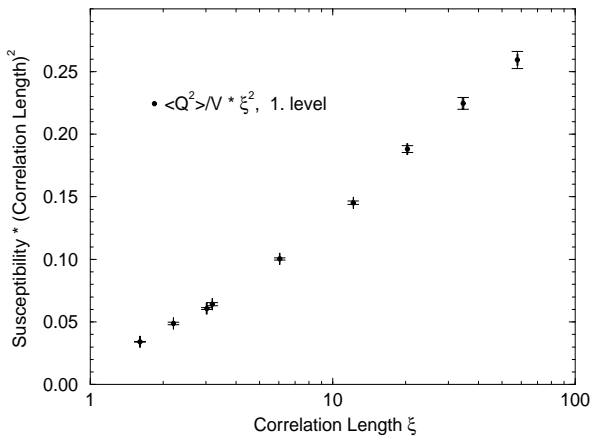

Figure 3. Results of Monte Carlo measurements of the topological susceptibility at different correlation lengths

## 5. Summary and Conclusions

We have defined a perfect action and a perfect topological charge with the properties:
– $\mathcal{A}^*$ has scale invariant instanton solutions,
– there are no topological defects
– and the topological charge $Q$ is integer.

Using these definitions we measured in the $d = 2$ O(3) non-linear $\sigma$-model the topological susceptibility to high precision but did not establish a scaling behaviour. The results suggest, that the topological susceptibility is not a physical quantity in this model. This statement is in agreement with semiclassical predictions [6] and recent numerical results [7], which suggest an ultraviolet dominance in the instanton size distribution.

We expect, however, that in $\mathbb{C}P^{N-1}$-models our method will give the correct topological susceptibility. Work on this subject is in progress.